\documentclass[aps,reprint,showpacs]{revtex4-1}
\usepackage{graphicx} 
\usepackage{tabularx}
\usepackage{dcolumn} 
\usepackage{color}
\usepackage{bm}
\usepackage{amsmath}
\usepackage{amssymb}

\newcommand{\bi}[1]{\ensuremath{\boldsymbol{#1}}} 
\newcommand{\nn}{\nonumber}

\begin{document}

\title{%
A Spin Triplet Superconductor UPt$_3$
}

\author{Yasumasa Tsutsumi} 
\affiliation{Department of Physics, Okayama University, 
Okayama 700-8530, Japan} 
\author{Kazushige Machida} 
\affiliation{Department of Physics, Okayama University, 
Okayama 700-8530, Japan} 
\author{Tetsuo Ohmi} 
\affiliation{Department of Physics, Kinki University, 
3-4-1 Kowakae, Higashi-Osaka, 577-8502, Japan} 
\author{Masa-aki Ozaki} 
\affiliation{Uji, 611-0002, Japan} 

\date{\today}

\begin{abstract}
Motivated by a recent angle-resolved thermal conductivity experiment that shows a twofold gap symmetry in the high-field and low-temperature C phase in the heavy-fermion superconductor UPt$_3$,
we group-theoretically identify the pairing functions as $E_{1u}$ with the $f$-wave character for all the three phases.
The pairing functions are consistent with the observation as well as with a variety of existing measurements.
By using a microscopic quasi-classical Eilenberger equation with the identified triplet pairing function under applied fields,
we performed detailed studies of the vortex structures for three phases,
including the vortex lattice symmetry, the local density of states, and the internal field distribution.
These quantities are directly measurable experimentally by SANS, STM/STS, and NMR, respectively.
It is found that, in the B phase of low $H$ and low $T$, the double-core vortex is stabilized over a singular vortex.
In the C phase, thermal conductivity data are analyzed to confirm the gap structure proposed.
We also give detailed comparisons of various proposed pair functions,
concluding that the present scenario of $E_{1u}$ with the $f$-wave, which is an analogue to the triplet planar state, 
is better than the $E_{2u}$ or $E_{1g}$ scenario.
Finally, we discuss the surface topological aspects of Majorana modes associated with the $E_{1u}^f$ state of planar like features.
\end{abstract}

\maketitle

\section{Introduction}

UPt$_3$~\cite{stewart:1984} belongs to the first generation of the family of 
heavy-fermion superconductors together with CeCu$_2$Si$_2$ and UBe$_{13}$
and has unique superconducting properties compared with them.
Immediately after the pioneering discovery~\cite{fisher:1989} of the double superconducting transition,
it is found~\cite{hasselbach:1989,bruls:1990,schenstrom:1989} that 
the phase diagram in the $H$ vs $T$ plane consists of the A, B, and C phases.
The A (C) phase is at a high (low) temperature and a low (high) field, while 
the B phase is at a low $T$ and a low $H$.
It is rather clear that the order parameter (OP)
must have multicomponents.
The main argument is centered on how to understand this phase diagram, or on
what OP can describe it in a consistent manner~\cite{sauls:1994,joynt:2002}.
Now the splitting of the superconducting transition temperatures 
$T_{c1}\cong 550$ mK and $T_{c2}\cong 500$ mK is generally understood owing to
a symmetry breaking field for an otherwise doubly degenerate pairing state.
The identification of this symmetry breaking field is still 
not settled yet, but it is considered to come from the antiferromagnetic (AF) 
ordering at $T_N=5$ K~\cite{joynt:1988,machida:1989,machida:1989b,machida:1991,ozaki:1992,hess:1989,tokuyasu:1990} 
or from the crystal lattice symmetry lowering
that occurs at higher temperatures~\cite{machida:1996}.

The remaining problem is identifying the OP symmetry. 
The central discussions are on the causes of OP degeneracy, that is, either 
the orbital part~\cite{joynt:1988,sauls:1994,hess:1989,tokuyasu:1990} 
or spin part of the OP~\cite{machida:1989,machida:1989b,machida:1991,ozaki:1992}.
The former scenario has a fundamental difficulty where the 
so-called gradient coupling term in the Ginzburg-Landau (GL) functional 
inevitably prevents the observed crossing of the two
transition lines starting from $T_{c1}$ and $T_{c2}$,
removing the C phase. Therefore, the orbital scenario needs 
the fine tuning of the underlying Fermi surface topology and
the detailed structure of the orbital function~\cite{sauls:1994,sauls:1994b}.
Among the various proposals, the $E_{2u}$ symmetry is regarded as 
the most possible candidate, where $\bi{d}(\bi{k})\propto\bi{z}(k_x^2-k_y^2+2ik_xk_y)k_z$.
This state is time-reversal-symmetry-broken and fourfold-symmetric in the
A and C phases. In the B phase, there exist one line node in the equator and
two point nodes in the poles.

On the other hand, the spin 
scenario~\cite{machida:1989,machida:1989b,machida:1991,ozaki:1992,ohmi:1993,machida:1993,machida:1995,ohmi:1996,machida:1999}
overcomes this difficulty,
but the difficulty to qualitatively estimate the spin-orbit (SO)
coupling remains because the spin scenario assumes a weak SO coupling 
in contrast to the strong SO coupling assumption adopted in the $E_{2u}$ 
scenario~\cite{sauls:1994,hess:1989,tokuyasu:1990,sauls:1994b,choi:1991,choi:1993}.
This controversy is resolved experimentally because the Knight shift~\cite{tou:1996,tou:1998} 
starts decreasing below $T_{c2}$ when $H\sim 2$ kG for $H\parallel c$.
This field $H_{\rm rot}$ corresponding to the rotation of the $d$-vector~\cite{tou:1998}
gives an estimate of the SO coupling strength in this system,
justifying the classification scheme from the weak SO coupling,
which is never attained in a strong-SO case, where the $d$-vector is 
strongly tied to the underlying crystalline axes via the orbital part in OP.

The basic requirements of the possible pairing state realized
in UPt$_3$ can be summarized as follows:
(1) The gap structure contains both horizontal line node(s) and point nodes
as evidenced by power law behaviors in various directionally dependent 
transport measurements, such as
thermal conductivity~\cite{lussier:1996b,suderow:1997} and ultrasound attenuation experiments~\cite{bishop:1984,muller:1986,shivaram:1986b,ellman:1996},
and also bulk measurements, such as specific heat~\cite{hasselbach:1989,brison:1994,ramirez:1995}, penetration depth~\cite{yaouanc:1998}, nuclear relaxation time~\cite{kohori:1988}, 
and magnetization~\cite{tenya:1996} experiments.
(2) As mentioned above, the detailed Knight shift experiment~\cite{tou:1998} shows a decrease in the
magnetic susceptibility below $T_{c2}$, depending on the field direction and its strength.
Thus, it is concluded that the $d$-vector contains the $\bi{b}$-component and $\bi{c}$-component
in the B phase for the hexagonal crystal. Upon increasing $H$ $(\parallel c)$, this
$\bi{c}$-component becomes the $\bi{a}$-component.
(3) Finally, according to the recent angle-resolved thermal conductivity measurement~\cite{machida:2012},
the twofold symmetric gap structure in the basal plane 
for the C phase, the full rotational symmetry in the B phase,
and the horizontal line nodes are found to be at the tropical position, not at the
equator of the Fermi sphere. 

In view of previous\cite{joynt:2002} and recent experiments\cite{machida:2012} we
come to a new stage to critically examine the proposed pairing states
belonging to the orbital scenario:
the singlet category $E_{1g}$\cite{park:1995} and the
triplet category $E_{2u}$\cite{hess:1989,tokuyasu:1990} in addition to the
so-called accidental degeneracy scenario\cite{joynt:1990,chen:1993} and also
belonging to the spin scenario~\cite{machida:1989,machida:1989b,machida:1991,ozaki:1992,ohmi:1993,machida:1993,machida:1995,ohmi:1996,machida:1999}.

Previously, we tentatively chose the orbital part from $E_{1u}$ representation~\cite{ohmi:1993,machida:1993,machida:1995,ohmi:1996}
where the OP is nonunitary and also that from
$E_{2u}$ representation~\cite{machida:1999} in our spin scenario among the classified pairing functions
in the absence~\cite{ozaki:1986,ozaki:1985} and presence of antiferromagnetism~\cite{ozaki:1989}. 
However, the precise orbital form is still to be determined.
In view of this finding we are now in the position to identify 
the precise pair functions for all A, B, and C phases, on the basis of the spin degeneracy scenario.

The paper is arranged as follows.
We first classify the pairing function group-theoretically and
examine the possible symmetry realized in connection with various
existing experimental data in \S II.
The formulation is given in \S III on the basis of the quasi-classical Eilenberger theory.
With the identified pair function, we 
investigate the vortex structures in the B phase, 
including the exotic vortex core structure, namely, the so-called
double-core structure associated with the multicomponent OP in \S IV.
The present study is a natural extension of our GL framework
for multiple OP~\cite{machida:1993b,fujita:1994,hirano:1995} to microscopic calculation based on the Eilenberger theory,
which provides quasiparticle (QP) structures in periodic vortex lattices.
A vortex structure is also studied for the A and C phases
in connection with the twofold gap function in \S V.
The final section is devoted to conclusions and future perspectives 
where detailed comparisons with other proposed pairing functions are made.
We use the two notations $(x,y,z)$ and $(a,b,c)$ to denote the spatial coordinates interchangeably.

\section{Classification of Possible Pairing States} 

In this section, we first enumerate possible pairing states group-theoretically~\cite{volovik:1985,ozaki:1986,ozaki:1985}.
The classified states are called inert phases, which are stable
under a small change in the system parameters and are different from
the previous classifications, where we only treat $p$-wave 
states~\cite{ozaki:1986,ozaki:1985}.
By the same procedure as those in refs.~\ref{ozaki1986} and \ref{ozaki1985},
we can easily extend it to $f$-wave pairing states
with an orbital angular momentum $l=3$.

Table I lists all the possible $f$-wave inert states and 
their little groups under the hexagonal symmetry $D_6$.
Here, the basis functions of the irreducible representations are
also given in Table I. 
As mentioned before, the required properties are all satisfied
by the planar state $\hat{\tau }_xl_1^{E_{1u}}+\hat{\tau }_yl_2^{E_{1u}}$ in $E_{1u}$.
Namely, in terms of the present context of UPt$_3$,
$(\bi{c}k_b+\bi{b}k_a)(5k_z^2-1)$,
where the unit vectors $\bi{a}$,
$\bi{b}$, and $\bi{c}$ in $D_6$ denote the $d$-vector components.
Note that this state is a natural extension of the planar state
$\hat{\tau }_xk_x+\hat{\tau }_yk_y$ in the $p$-wave state that is realized in thin 
films of the superfluid $^3$He B-phase.

\begin{table*}[t]
\begin{center}
\caption{$f$-Wave inert phases and their little groups in $D_6$ system.}
{\tabcolsep=0.5mm
\begin{tabular}{lllll}
\hline\hline
rep. & state & order parameter & basis of irr.~rep. & little group\\\hline
$A_{2u}$ & $A_{2u}$-polar & $\hat{\tau }_zl^{A_{2u}}$ & $l^{A_{2u}}\!=\!2k_z^3\!-\!3(k_x^2\!+\!k_y^2)k_z$ & $(1\!+\!C_{21}'u_{2x})(1\!+\!C_{21}'\tilde{\pi })\{\bi{C}_6\!\times\! \bi{A}(\bi{e}_z)\!\times\! T\}$ \\
 & $A_{2u}$-$\beta$ & $(\hat{\tau }_x\!+\!i\hat{\tau }_y)l^{A_{2u}}$ & & $(1\!+\!tu_{2x})(1\!+\!C_{21}'u_{2z})\{\bi{C}_6\!\times\! \tilde{\bi{A}}(\bi{e}_z)\}$ \\\hline
$B_{1u}$ & $B_{1u}$-polar & $\hat{\tau }_zl^{B_{1u}}$ & $l^{B_{1u}}\!=\!k_x^3\!-\!3k_y^2k_x$ & $(1\!+\!C_2u_{2x})(1\!+\!C_{21}''\tilde{\pi })\{\bi{D}_3'\!\times\! \bi{A}(\bi{e}_z)\!\times\! T\}$ \\
 & $B_{1u}$-$\beta$ & $(\hat{\tau }_x\!+\!i\hat{\tau }_y)l^{B_{1u}}$ & & $(1\!+\!tu_{2x})(1\!+\!C_2u_{2z})\{\bi{D}_3'\!\times\! \tilde{\bi{A}}(\bi{e}_z)\}$ \\\hline
$B_{2u}$ & $B_{2u}$-polar & $\hat{\tau }_zl^{B_{2u}}$ & $l^{B_{2u}}\!=\!3k_yk_x^2\!-\!k_y^3$ & $(1\!+\!C_2u_{2x})(1\!+\!C_{21}'\tilde{\pi })\{\bi{D}_3''\!\times\! \bi{A}(\bi{e}_z)\!\times\! T\}$ \\
 & $B_{2u}$-$\beta$ & $(\hat{\tau }_x\!+\!i\hat{\tau }_y)l^{B_{2u}}$ & & $(1\!+\!tu_{2x})(1\!+\!C_{21}'u_{2z})\{\bi{D}_3''\!\times\! \tilde{\bi{A}}(\bi{e}_z)\}$ \\\hline
$E_{1u}$ & $E_{1u}$-planar & $\hat{\tau }_xl_1^{E_{1u}}\!+\!\hat{\tau }_yl_2^{E_{1u}}$ & $l_1^{E_{1u}}\!=\!(5k_z^2\!-\!1)k_x$ & $(1\!+\!C_{21}'u_{2y}\tilde{\pi })\{_{II}\bi{D}_6\!\times\! T\}$ \\
 & $E_{1u}$-polar$_1$ & $\hat{\tau }_zl_1^{E_{1u}}$ & $l_2^{E_{1u}}\!=\!(5k_z^2\!-\!1)k_y$ & $(1\!+\!C_{2z}\tilde{\pi })(1\!+\!C_{2z}u_{2x})\{\bi{C}_{21}'\!\times\! \bi{A}(\bi{e}_z)\!\times\! T\}$ \\
 & $E_{1u}$-polar$_2$ & $\hat{\tau }_zl_2^{E_{1u}}$ & & $(1\!+\!C_{2z}\tilde{\pi })(1\!+\!C_{2z}u_{2x})\{\bi{C}_{21}''\!\times\! \bi{A}(\bi{e}_z)\!\times\! T\}$ \\
 & $E_{1u}$-bipolar & $\hat{\tau }_xl_1^{E_{1u}}\!+\!i\hat{\tau }_yl_2^{E_{1u}}$ & & $(1\!+\!tu_{2x})(1\!+\!C_{2z}\tilde{\pi })_{II}\bi{D}_2$ \\
 & $E_{1u}$-axial & $\hat{\tau }_z(l_1^{E_{1u}}\!+\!il_2^{E_{1u}})$ & & $(1\!+\!tC_{21}')(1\!+\!u_{2x}\tilde{\pi })\{\tilde{\bi{C}}_6\!\times\! \bi{A}(\bi{e}_z)\}$ \\
 & $E_{1u}$-$\beta_1$ & $(\hat{\tau }_x\!+\!i\hat{\tau }_y)l_1^{E_{1u}}$ & & $(1\!+\!tu_{2x})(1\!+\!C_{2z}u_{2z})\{\bi{C}_{21}'\!\times\! \tilde{\bi{A}}(\bi{e}_z)\}$ \\
 & $E_{1u}$-$\beta_2$ & $(\hat{\tau }_x\!+\!i\hat{\tau }_y)l_2^{E_{1u}}$ & & $(1\!+\!tu_{2x})(1\!+\!C_{2z}u_{2z})\{\bi{C}_{21}''\!\times\! \tilde{\bi{A}}(\bi{e}_z)\}$ \\
 & $E_{1u}$-$\gamma$ & $(\hat{\tau }_x\!+\!i\hat{\tau }_y)(l_1^{E_{1u}}\!+\!il_2^{E_{1u}})$ & & $(1\!+\!tC_{21}'u_{2x})\{\tilde{\bi{C}}_6\!\times\! \tilde{\bi{A}}(\bi{e}_z)\}$ \\\hline
$E_{2u}$ & $E_{2u}$-planar & $\hat{\tau }_xl_1^{E_2u}\!+\!\hat{\tau }_yl_2^{E_{2u}}$ & $l_1^{E_{2u}}\!=\!2k_zk_xk_y$ & $(1\!+\!C_{21}'u_{2x})(1\!+\!u_{2z}\tilde{\pi })\{_{II}\bi{C}_6^2\!\times\! T\}$ \\
 & $E_{2u}$-polar$_1$ & $\hat{\tau }_zl_1^{E_{2u}}$ & $l_2^{E_{2u}}\!=\!-k_z(k_x^2\!-\!k_y^2)$ & $(1\!+\!u_{2x}\tilde{\pi })\{\bi{D}_2\!\times\! \bi{A}(\bi{e}_z)\!\times\! T\}$ \\
 & $E_{2u}$-polar$_2$ & $\hat{\tau }_zl_2^{E_{2u}}$ & & $(1\!+\!u_{2x}\tilde{\pi })(1\!+\!C_{21}'u_{2x})\{\bi{C}_2\!\times\! \bi{A}(\bi{e}_z)\!\times\! T\}$ \\
 & $E_{2u}$-bipolar & $\hat{\tau }_xl_1^{E_{2u}}\!+\!i\hat{\tau }_yl_2^{E_{2u}}$ & & $(1\!+\!tu_{2x})(1\!+\!C_{21}'u_{2x})\bi{C}_2$ \\
 & $E_{2u}$-axial & $\hat{\tau }_z(l_1^{E_{2u}}\!+\!il_2^{E_{2u}})$ & & $(1\!+\!tC_{21}')\{\tilde{\bi{C}}_6^2\!\times\! \bi{A}(\bi{e}_z)\}$ \\
 & $E_{2u}$-$\beta_1$ & $(\hat{\tau }_x\!+\!i\hat{\tau }_y)l_1^{E_{2u}}$ & & $(1\!+\!tu_{2x})\{\bi{D}_2\!\times\! \tilde{\bi{A}}(\bi{e}_z)\}$ \\
 & $E_{2u}$-$\beta_2$ & $(\hat{\tau }_x\!+\!i\hat{\tau }_y)l_2^{E_{2u}}$ & & $(1\!+\!tu_{2x})(1\!+\!C_{21}'u_{2z})\{\bi{C}_2\!\times\! \tilde{\bi{A}}(\bi{e}_z)\}$ \\
 & $E_{2u}$-$\gamma$ & $(\hat{\tau }_x\!+\!i\hat{\tau }_y)(l_1^{E_{2u}}\!+\!il_2^{E_{2u}})$ & & $(1\!+\!tC_{21}'u_{2x})\{\tilde{\bi{C}}_6^2\!\times\! \tilde{\bi{A}}(\bi{e}_z)\}$ \\\hline
\multicolumn{5}{l}{
$\tilde{\bi{A}}(\bi{e}_z)\!=\!\{u(\bi{e}_z,\theta)\tilde{\theta }|0\le\theta\le 2\pi\}$,
$\bi{C}_2\!=\!\{E\!+\!C_{2z}\}$,
$\bi{C}_{21}'\!=\!\{E\!+\!C_{21}'\}$,
$\bi{D}_3'\!=\!\{\bi{C}_3\!+\!C_{21}'\bi{C}_3\}$,
}\\
\multicolumn{5}{l}{
$\bi{D}_3''\!=\!\{\bi{C}_3\!+\!C_{21}''\bi{C}_3\}$,
$_{II}\bi{D}_2\!=\!\{1,C_{2z}u_{2z},C_{21}'u_{2x},C_{21}''u_{2y}\}$,
$\tilde{\bi{C}}_6\!=\!\{C_6^j(\frac{2\pi j}{6}),j\!=\!0,1,\cdots,5\}$,
}\\
\multicolumn{5}{l}{
$\tilde{\bi{C}}_6^2\!=\!\{C_6^j(\frac{2\cdot 2\pi j}{6}),j\!=\!0,1,\cdots,5\}$,
$_{II}\bi{C}_6^2\!=\!\{C_6^ju_6^{2j},j\!=\!0,1,\cdots,5\}$,
$\hat{\tau }_{\mu }\!=\!i\hat{\sigma }_{\mu }\hat{\sigma }_y$ $(\mu\!=\!x,y,z)$.
}\\
\hline\hline 
\end{tabular}}
\end{center}
\end{table*}

\begin{figure}
\begin{center}
\includegraphics[width=8.5cm]{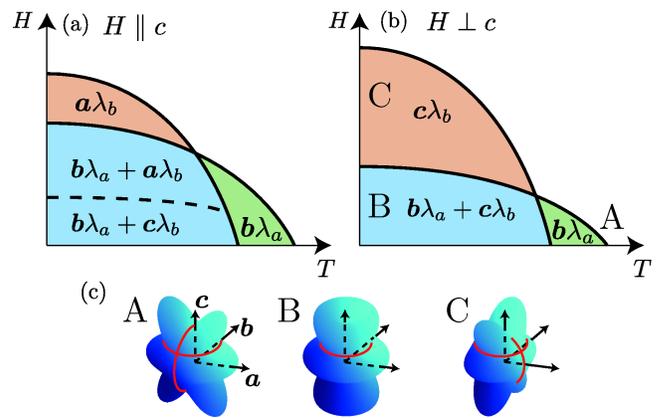}
\end{center}
\caption{(Color online) 
Schematic phase diagrams under $H\parallel c$ (a) and $H\perp c$ (b).
The orbital states are $\lambda_a(\bi{k})=\sqrt{21/8}k_a(5k_c^2-1)$ and $\lambda_b(\bi{k})=\sqrt{21/8}k_b(5k_c^2-1)$.
(c) Gap functions of the A, B, and C phases.
} 
\label{phase}
\end{figure}

\subsection{Phase assignment and gap structures} 

\begin{figure}
\begin{center}
\includegraphics[width=7cm]{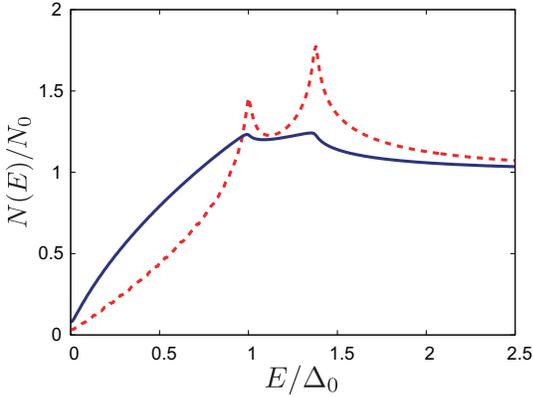}
\end{center}
\caption{(Color online) 
DOS's in the bulk for A and C phases $|\Delta(\bi{k})|=\Delta_0|k_x(5k_z^2-1)|$ (solid line)
and B phase $|\Delta(\bi{k})|=\Delta_0\sqrt{k_x^2+k_y^2}|5k_z^2-1|$ (dashed line).
} 
\label{bulkDOS}
\end{figure}

This $E_{1u}$ state is assigned to each phase as
the B phase $(\bi{c}k_b+\bi{b}k_a)(5k_c^2-1)$, the A
phase $\bi{b}k_a(5k_c^2-1)$, and the C phase $\bi{c}k_b(5k_c^2-1)$, as shown in Fig.~\ref{phase}.
Thus, the B phase is characterized by two horizontal line nodes at
$\cos^2\theta=1/5$, or $\theta=63.4^{\circ}$ and $116.6^{\circ}$
and two point nodes at the poles [see Fig.~\ref{phase}(c)].
The C (A) phase is the vertical line node at $k_b=0$ ($k_a=0$) in addition to
the two horizontal line nodes [see Fig.~\ref{phase}(c)]. Thus, it naturally explains
the twofold thermal conductivity oscillation when $H$ rotates within the basal plane.
Note also that this gap structure with two line nodes and two point nodes in the B
phase is consistent with the various transport measurements and bulk measurements
mentioned above.
In Fig.~\ref{bulkDOS}, we show the densities of states (DOS's) $N(E)$ for the three phases.

\subsection{Unitary versus nonunitary} 

It is shown in Table I that we list up the bipolar state
given by $\hat{\tau }_xl_1^{E_{1u}}+i\hat{\tau }_yl_2^{E_{1u}}$ that is nonunitary 
and breaks the time reversal symmetry and that can also explain the phase diagram
in $H$ vs $T$. However, it is not consistent with the recent $\mu$SR experiment~\cite{reotier:1995}
that negates the earlier claim that the time reversal symmetry is broken~\cite{luke:1993}.
The gap structure in this bipolar state is characterized by $|k_x\pm k_y||5k_z^2-1|$ 
in the B phase with fourfold symmetry. This contradicts 
the results of a thermal conductivity experiment~\cite{machida:2012} that indicated the rotational
symmetry in the B phase.

\subsection{Other classified states and strong SO state} 

The remaining states among the classified inert phase, i.e.,
$A_2$, $B_1$, $B_2$, $E_1$, and $E_2$, are not accepted for
the following reasons:
They do not provide the double transition ($A_2$, $B_1$, and $B_2$)
or explain the twofold symmetry in the C phase ($E_2$).
The other $E_{1u}$ states in Table I are not appropriate except for the planar state,
namely, the polar$_1$, polar$_2$, axial, $\beta_1$, $\beta_2$, and $\gamma$,
either because they do not give the double transition (polar$_1$ and polar$_2$), or
because they fail to give the appropriate gap structure required (axial, $\beta_1$, $\beta_2$, and $\gamma$). 

The $E_{2u}$ $\bi{c}(k_a^2-k_b^2+2ik_ak_b)k_c$ classified
in the strong-SO case fails to explain the Knight shift experiment~\cite{tou:1996,tou:1998}
and the twofold symmetry in the C phase~\cite{machida:2012}. Then we are left with only the
$E_{1u}$ planar state with the $f$-wave character mentioned above.
Note also that, since $E_{1u}$ with the $p$-wave character $\hat{\tau }_xk_x+i\hat{\tau }_yk_y$
has no line node, it has been excluded as a candidate from the outset.

\begin{widetext}
\section{Quasi-Classical Eilenberger Theory}

We start with the quasi-classical spinful Eilenberger equation~\cite{eilenberger:1968,schopohl:1980,serene:1983,fogelstrom:1995,sauls:2009}.
The quasi-classical Green's function $\widehat{g}(\bi{k},\bi{r},\omega_n)$ is calculated using the Eilenberger equation
\begin{align}
-i\hbar\bi{v}(\bi{k})\cdot\bi{\nabla }\widehat{g}(\bi{k},\bi{r},\omega_n) 
= \left[
\begin{pmatrix}
\left[i\omega_n+(e/c)\bi{v}(\bi{k})\cdot\bi{A}(\bi{r})\right]\hat{1} & -\hat{\Delta }(\bi{k},\bi{r}) \\
\hat{\Delta }(\bi{k},\bi{r})^{\dagger } & -\left[i\omega_n+(e/c)\bi{v}(\bi{k})\cdot\bi{A}(\bi{r})\right]\hat{1}
\end{pmatrix}
,\widehat{g}(\bi{k},\bi{r},\omega_n) \right],
\label{Eilenberger eq}
\end{align}
\end{widetext}
where the  ordinary hat indicates the 2 $\times$ 2 matrix in spin space 
and the wide hat indicates the 4 $\times$ 4 matrix in particle-hole and spin spaces.
The quasi-classical Green's function is described in particle-hole space by
\begin{align}
\widehat{g}(\bi{k},\bi{r},\omega_n) = -i\pi
\begin{pmatrix}
\hat{g}(\bi{k},\bi{r},\omega_n) & i\hat{f}(\bi{k},\bi{r},\omega_n) \\
-i\underline{\hat{f}}(\bi{k},\bi{r},\omega_n) & -\underline{\hat{g}}(\bi{k},\bi{r},\omega_n)
\end{pmatrix},
\end{align}
with the direction of the relative momentum of a Cooper pair $\bi{k}$, the center-of-mass coordinate of the Cooper pair $\bi{r}$,
and the Matsubara frequency $\omega_n=(2n+1)\pi k_B T$.
The quasi-classical Green's function satisfies the normalization condition $\widehat{g}^2=-\pi^2\widehat{1}$.
The Fermi velocity is assumed as $\bi{v}(\bi{k})=v_F\bi{k}$ on a three-dimensional Fermi sphere.
In the symmetric gauge, the vector potential $\bi{A}(\bi{r})=(\bar{\bi{B}}\times\bi{r})/2+\bi{a}(\bi{r})$,
where $\bar{\bi{B}}=(0,0,\bar{B})$ is a uniform flux density and $\bi{a}(\bi{r})$ is related to the internal field $\bi{B}(\bi{r})=\bar{\bi{B}}+\nabla\times\bi{a}(\bi{r})$.
The unit cell of the vortex lattice is given by $\bi{r}=s_1(\bi{u}_1-\bi{u}_2)+s_2\bi{u}_2$ with $-0.5\le s_i\le 0.5$ $(i=1,2)$,
$\bi{u}_1=(a_x,0,0)$, and $\bi{u}_2=(a_x/2,a_y,0)$.
In this coordinate, a hexagonal lattice is described by $a_y/a_x=\sqrt{3}/2$ or $1/(2\sqrt{3})$.

The spin triplet order parameter is defined by
\begin{align}
\hat{\Delta }(\bi{k},\bi{r})=i\bi{d}(\bi{k},\bi{r})\cdot\hat{\bi{\sigma }}\hat{\sigma_y},
\end{align}
with
\begin{align}
\bi{d}(\bi{k},\bi{r})=\bi{a}\Delta_a(\bi{r})\phi_a(\bi{k})+\bi{b}\Delta_b(\bi{r})\phi_b(\bi{k})+\bi{c}\Delta_c(\bi{r})\phi_c(\bi{k}),
\end{align}
where $\hat{\bi{\sigma }}$ is the Pauli matrix.
The self-consistent condition for $\Delta_i(\bi{r})$ is given as
\begin{multline}
\hat{\Delta }(\bi{k},\bi{r}) = N_0\pi k_BT \\
\times\sum_{0<\omega_n \le \omega_c}\left\langle V(\bi{k}, \bi{k}') \left[\hat{f}(\bi{k}',\bi{r},\omega_n)+\underline{\hat{f}}^{\dagger }(\bi{k}',\bi{r},\omega_n)\right]\right\rangle_{\bi{k}'},
\label{order parameter}
\end{multline}
where $N_0$ is the DOS in the normal state,
$\omega_c$ is the cutoff energy setting $\omega_c=20\pi k_B T_c$ with the transition temperature $T_c$,
and $\langle\cdots\rangle_{\bi{k}}$ indicates the Fermi surface average.
We neglect the splitting of $T_c$ because it is appropriate at low temperatures even in the B phase.
The pairing interaction $V(\bi{k}, \bi{k}')=g\phi(\bi{k})\phi^*(\bi{k}')$, where $g$ is a coupling constant.
The pairing functions $\phi(\bi{k})$ and $\phi_i(\bi{k})$ are chosen for each phase in UPt$_3$ appropriately.
In our calculation, we use the relation
\begin{align}
\frac{1}{gN_0}=\ln \frac{T}{T_c}+2\pi k_BT\sum_{0<\omega_n \le \omega_c}\frac{1}{\omega_n}.
\end{align}
The vector potential for the internal magnetic field $\bi{A}(\bi{r})$ is also self-consistently determined by
\begin{align}
&\nabla\times[\nabla\times\bi{A}(\bi{r})]\nn\\
=&8\pi\frac{e}{c}N_02\pi k_BT\sum_{0<\omega_n\le\omega_c}
\langle \bi{v}(\bi{k}) \ {\rm Im} \left[ g_0(\bi{k},\bi{r}, \omega_n) \right] \rangle_{\bi{k}},
\label{vector potential}
\end{align}
where $g_0$ is a component of the quasi-classical Green's function $\hat{g}$ in spin space, namely,
\begin{align}
\hat{g} =
\begin{pmatrix}
g_0+g_z & g_x-ig_y \\
g_x+ig_y & g_0-g_z
\end{pmatrix}. \nn
\end{align}

We solve eq.~\eqref{Eilenberger eq} and eqs.~\eqref{order parameter} and \eqref{vector potential} alternately,
and obtain self-consistent solutions, under a given unit cell of the vortex lattice.
The unit cell is divided into $41\times 41$ mesh points,
where we obtain the quasi-classical Green's functions, $\Delta_i(\bi{r})$, and $\bi{A}(\bi{r})$.
When we solve eq.~\eqref{Eilenberger eq} by the Riccati method~\cite{nagato:1993,schopohl:1995},
we estimate $\Delta_i(\bi{r})$, and $\bi{A}(\bi{r})$ at arbitrary positions by the interpolation from their values at the mesh points,
and by the periodic boundary condition of the unit cell including the phase factor due to the magnetic field~\cite{ichioka:1997,ichioka:1999a,ichioka:1999b}.
In the numerical calculation, we use the units $R_0=\hbar v_F/(2\pi k_BT_c)$, $B_0=\hbar c/(2|e|R_0^2)$, and $E_0=\pi k_BT_c$
for the length, magnetic field, and energy, respectively.
By the dimensionless expression, eq.~\eqref{vector potential} is rewritten as
\begin{align}
&\frac{R_0}{B_0}\nabla\times[\nabla\times\bi{A}(\bi{r})]\nn\\
=&-\frac{1}{\kappa^2}\frac{2T}{T_c}\sum_{0<\omega_n\le\omega_c}
\langle \bi{k} \ {\rm Im} \left[ g_0(\bi{k},\bi{r}, \omega_n) \right] \rangle_{\bi{k}},
\end{align}
where $\kappa=B_0/(E_0\sqrt{8\pi N_0})=\sqrt{7\zeta(3)/18}\kappa_{\rm GL}$.
We use a large GL parameter $\kappa_{\rm GL}=60$ owing to UPt$_3$.

By using the self-consistent solutions, free energy density is calculated using Luttinger-Ward thermodynamic potential~\cite{luttinger:1960} as
\begin{widetext}
\begin{align}
\delta\Omega=&N_0\frac{1}{gN_0}\left\langle\left\langle\frac{1}{2}{\rm Tr}\hat{\Delta }(\bi{k},\bi{r})\hat{\Delta }^{\dagger }(\bi{k},\bi{r})\right\rangle_{\bi{k}}\right\rangle_{\bi{r}}
+N_0E_0^2\kappa^2\left\langle\left(\frac{\nabla\times\bi{A}(\bi{r})-\bar{\bi{B}}}{B_0}\right)^2\right\rangle_{\bi{r}}\nn\\
-&N_0\int_0^1d\lambda\left\langle\pi k_BT\sum_{0<\omega_n\le\omega_c}\left\langle{\rm Re}\left[{\rm Tr}\left\{\hat{\Delta }^{\dagger }(\bi{k},\bi{r})\left(\hat{f}_{\lambda }(\bi{k},\bi{r},\omega_n)\!+\!\underline{\hat{f}}_{\lambda }^{\dagger }(\bi{k},\bi{r},\omega_n)\right)\right\}\right]\right\rangle_{\bi{k}}\right\rangle_{\bi{r}},
\label{Luttinger-Ward}
\end{align}
\end{widetext}
where $\langle\cdots\rangle_{\bi{r}}$ indicates the spatial average.
The auxiliary functions $\hat{f}_{\lambda }$ and $\underline{\hat{f}}_{\lambda }$ are obtained 
by the substitution of $\lambda\hat{\Delta }$ for $\hat{\Delta }$ in eq.~\eqref{Eilenberger eq}.
This thermodynamic potential is relevant under large GL parameters and low magnetic fields
because the replacement of the vector potential is not carried out.

DOS for the energy $E$ is given by
\begin{align}
\bar{N}(E)=&\langle N(\bi{r},E)\rangle_{\bi{r}}\nn\\
=&\left\langle N_0 \left\langle {\rm Re} \left[g_0(\bi{k},\bi{r}, \omega_n)|_{i\omega_n \rightarrow E+i\eta}\right] \right\rangle_{\bi{k}}\right\rangle_{\bi{r}},
\end{align}
where $\eta$ is a positive infinitesimal constant and $N(\bi{r},E)$ is the local density of states (LDOS).
Typically, we use $\eta=0.01\pi k_BT_c$.
To obtain $g_0(\bi{k},\bi{r}, \omega_n)|_{i\omega_n \rightarrow E+i\eta}$, 
we solve eq.~\eqref{Eilenberger eq} with $\eta -iE$ instead of $\omega_n$ under the pair potential and vector potential obtained self-consistently.

\section{B Phase}

In the B phase, we take the pairing functions as $\phi=\sqrt{21/8}(k_a+k_b)(5k_c^2-1)$, $\phi_b=\sqrt{21/8}k_a(5k_c^2-1)$,
and $\phi_a=\phi_c=\sqrt{21/8}k_b(5k_c^2-1)$,
where one component of the $d$-vector is directed toward the $b$-axis and the other can rotate in the $ac$-plane.

\subsection{Double-core vortex lattice}

The pairing function in the B phase is similar to that in the superfluid $^3$He B-phase,
namely, $\bi{d}(\bi{k})\propto\bi{x}k_x+\bi{y}k_y+\bi{z}k_z$~\cite{vollhardt:book}.
Owing to the analogy with the $^3$He B-phase,
there is the possibility that the unconventional double-core vortex~\cite{thuneberg:1986b,thuneberg:1987} and $v$ vortex~\cite{salomaa:1983,salomaa:1985b} with a chiral core are stabilized against the conventional singular vortex.
In fact, double-core vortex and $v$ vortex are stabilized in the low- and high-pressure regions, respectively, in the $^3$He B-phase~\cite{thuneberg:1986b,thuneberg:1987}.
Under our pairing function in the UPt$_3$ B phase, the $v$ (chiral core) vortex lattice is not stabilized self-consistently; 
however, there are two types of self-consistent vortex lattice,
namely, the hexagonal singular vortex lattice and the double-core vortex lattice.

\begin{figure}
\begin{center}
\includegraphics[width=8.5cm]{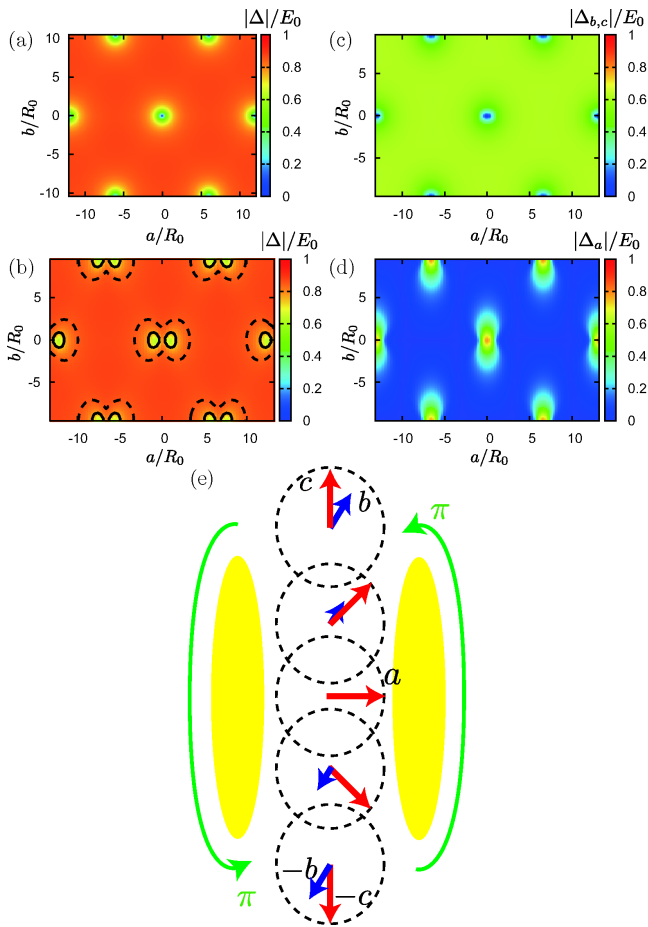}
\end{center}
\caption{(Color online) 
Spatial variations in the pair potential at $T=0.2T_c$ and $\bar{B}=0.05B_0$ for the hexagonal singular vortex lattice (a)
and the double-core vortex lattice with $\alpha\approx 55^{\circ }$ (b)-(d).
(a) Amplitude of the total pair potential $|\Delta(\bi{r})|$,
(b) amplitude of the total pair potential $|\Delta(\bi{r})|$ with the contour lines on $0.75E_0$ (solid lines) and $0.85E_0$ (dashed lines),
(c) amplitude of the bulk components $|\Delta_b(\bi{r})|=|\Delta_c(\bi{r})|$,
(d) amplitude of the compensating component at the vortex cores $|\Delta_a(\bi{r})|$, and
(e) schematic spin structure of the double-core vortex around the core.
} 
\label{OP-B}
\end{figure}

At $T=0.2T_c$ under $\bar{B}=0.05B_0$ to the $c$-axis, a spatial variation of the pair potential amplitude for the singular vortex lattice and the double-core vortex lattice is shown in Figs.~\ref{OP-B}(a) and \ref{OP-B}(b)-\ref{OP-B}(d), respectively,
where the total pair potential is defined by $|\Delta(\bi{r})|\equiv\sqrt{\langle{\rm Tr}[\hat{\Delta }(\bi{k},\bi{r})^{\dagger }\hat{\Delta }(\bi{k},\bi{r})]/2\rangle_{\bi{k}}}$.
Since the pair potential is axisymmetric for the $c$-axis in the B phase, 
conventional singular vortices form a perfect hexagonal lattice [Fig.~\ref{OP-B}(a)].
By contrast, a double-core vortex spontaneously breaks the axisymmetry.
A schematic structure of the double-core vortex by the $d$-vector is shown in Fig.~\ref{OP-B}(e).
The OP in the bulk is depicted by the blue (black) and red (gray) arrows, 
which indicate components of the $d$-vector with the orbital states $\lambda_a=\sqrt{21/8}k_a(5k_c^2-1)$ and $\lambda_b=\sqrt{21/8}k_b(5k_c^2-1)$, respectively.
Along the $b$-axis across the vortex center, the red (gray) arrow rotates in the $ac$-plane from the $c$-direction far from the vortex to the $-c$-direction on the opposite side of the vortex via the $a$-direction at the vortex center.
On the other hand, the blue (black) arrow directed toward the $b$-direction shortens as it approaches the vortex center and finally vanishes at the vortex center; 
then, across the vortex center, the arrow lengthens toward the $-b$-direction up to the initial length.
Thus, since the $d$-vector can be modulated continuously across the vortex center,
there is no singularity where the total amplitude vanishes.
Instead of a singularity, a double core with a small amplitude $\approx 0.6E_0$ exists, as shown by contour lines in Fig.~\ref{OP-B}(b).
This double core has a phase winding $\pi$ the same as that in the half-quantum vortex~\cite{volovik:1976}.

By the spontaneously broken axisymmetry, the double-core vortex lattice is distorted.
The stable ratio between the height and base of the triangular lattice is $a_y/a_x=\sqrt{3}/2.4$,
namely, a base angle $\alpha\equiv\tan^{-1}(2a_y/a_x)\approx 55^{\circ }$.
Each component of the double-core vortex lattice is also shown in Figs.~\ref{OP-B}(c) and \ref{OP-B}(d)
for the amplitude of the bulk components $|\Delta_b(\bi{r})|=|\Delta_c(\bi{r})|$
and that of the compensating component at the vortex cores $|\Delta_a(\bi{r})|$, respectively.
The vortex core in the bulk component is slightly elliptic with a line of apsides along the $a$-axis
and the compensating component is enlarged along the $b$-axis.
Since the vortex lattice tends to prevent the overlap of the vortex cores and that of the compensating component,
the stable structure is fixed by the competition between them.

At low temperatures and low magnetic fields, the double-core vortex is more stable than the singular vortex.
At high temperatures, however, the double-core vortex is unstable against the singular vortex
because the compensating component tends to connect with the neighbor vortices along the $b$-axis by the extension of the coherence length.
Similarly, at high magnetic fields, since the distance between the vortex centers becomes shorter, the double-core vortex is unstable.
The Pauli-paramagnetic effect, which rotates the $d$-vector under $H>H_{\rm rot}$, is neglected in this calculation.
Since the compensating component of the $d$-vector for $\bi{b}\lambda_a+\bi{a}\lambda_b$ has to be directed toward the $c$-axis,
the double-core vortex is unstable at high magnetic fields $H>H_{\rm rot}$, also by the Pauli-paramagnetic effect.

By the measurement of small-angle neutron scattering (SANS),
a perfect hexagonal lattice with $\alpha=60^{\circ }$ is observed in the B phase~\cite{huxley:2000}.
Thus, the observed vortices are conventional singular vortices.
Since this experiment is carried out at a magnetic field $H\approx H_{\rm rot}$,
the double-core vortex may be unstable, by the Pauli-paramagnetic effect.

\subsection{Local density of states}

\begin{figure}
\begin{center}
\includegraphics[width=8cm]{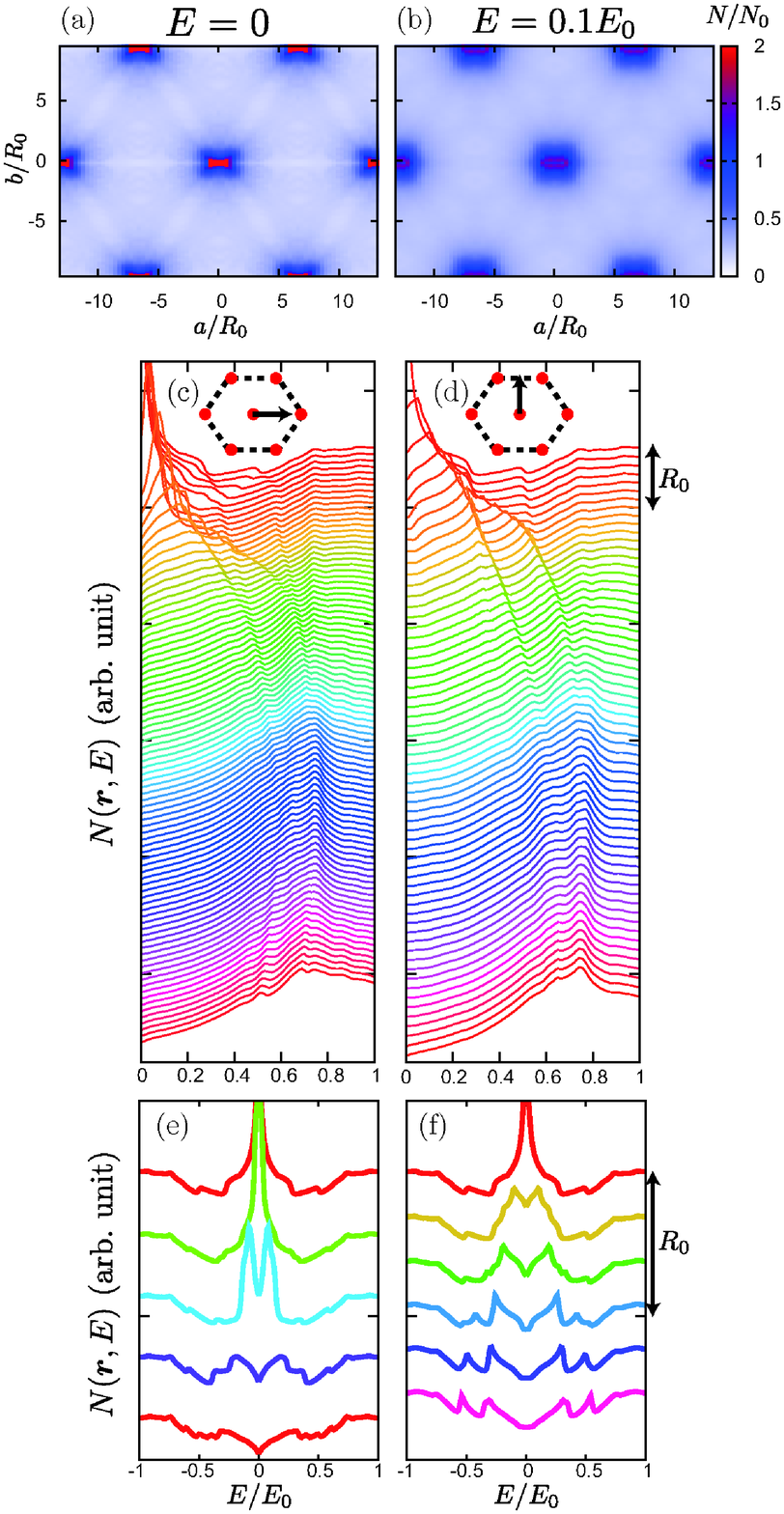}
\end{center}
\caption{(Color online) 
LDOS at $T=0.2T_c$ and $\bar{B}=0.05B_0$ for the double-core vortex lattice.
Spatial variations in the LDOS at $E=0$ (a) and $E=0.1E_0$ (b).
Spectral evolutions of the LDOS from the vortex center (top) along the $a$-axis (c) and $b$-axis (d).
Details of these evolutions near the vortex center are shown in (e) and (f), respectively.
} 
\label{LDOS-d}
\end{figure}

\begin{figure}
\begin{center}
\includegraphics[width=7cm]{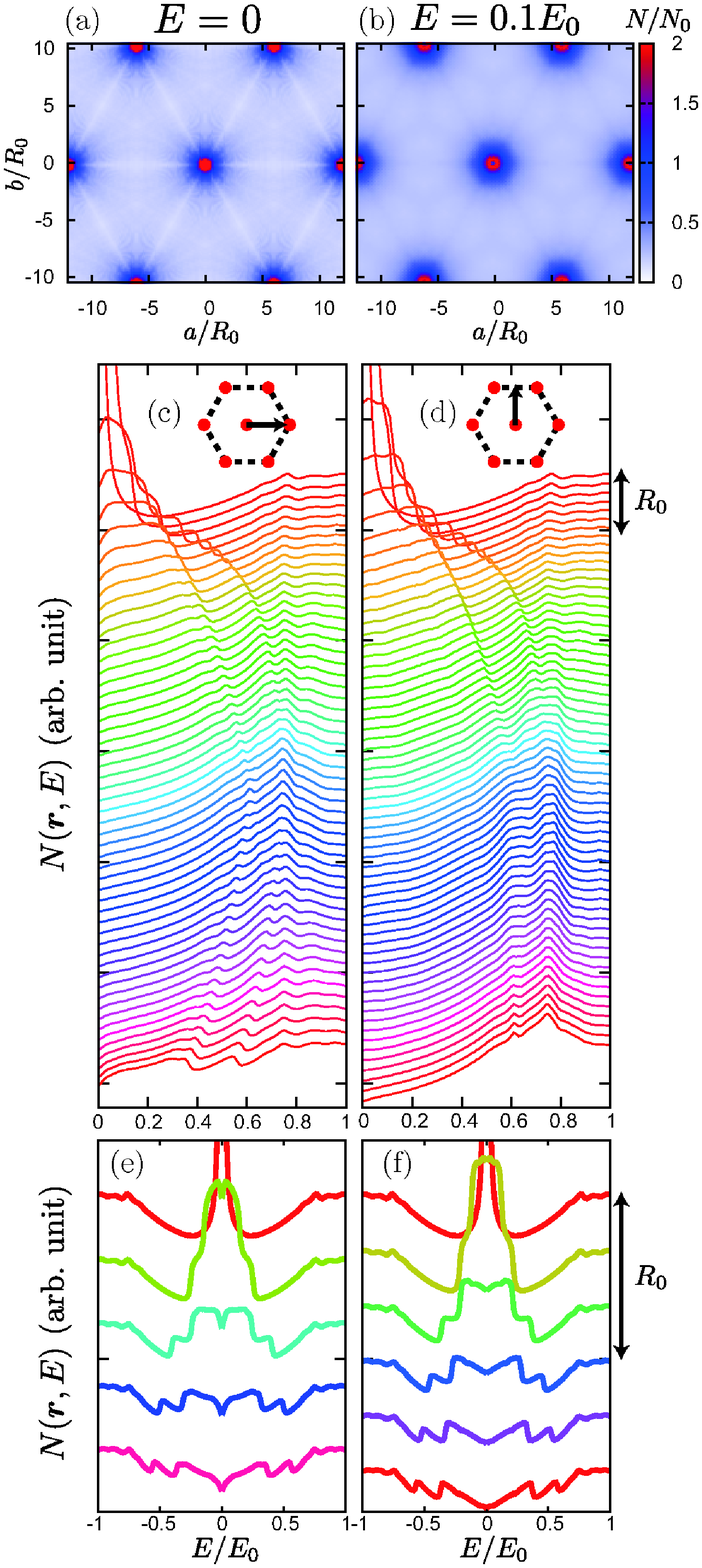}
\end{center}
\caption{(Color online) 
LDOS at $T=0.2T_c$ and $\bar{B}=0.05B_0$ for the singular vortex lattice.
Spatial variations in the LDOS at $E=0$ (a) and $E=0.1E_0$ (b).
Spectral evolutions of the LDOS from the vortex core (top) along the $a$-axis (c) and $b$-axis (d).
Details of these evolutions near the vortex core are shown in (e) and (f), respectively.
} 
\label{LDOS-s}
\end{figure}

There are clear differences in the LDOS between the double-core vortex lattice and the singular vortex lattice,
which can be directly measured by scanning tunneling microscopy/spectroscopy (STM/STS).
The LDOS's for the double-core vortex lattice and the singular vortex lattice are shown in Figs.~\ref{LDOS-d} and \ref{LDOS-s}, respectively.
The zero-energy peak is expanded to the region between the double core [Fig.~\ref{LDOS-d}(a)]
because the local OP $\bi{a}\lambda_b$ in this region has a line node in the $ac$-plane.
Besides, there is an elliptic peak extending toward the $a$-axis in the LDOS at $E=0.1E_0$ [Fig.~\ref{LDOS-d}(b)].
By contrast, the singular vortex has an isotropic peak in the LDOS at $E=0$ and $E=0.1E_0$ [Figs.~\ref{LDOS-s}(a) and \ref{LDOS-s}(b)].
The spectral evolutions of the LDOS near the vortex are also different between the double-core vortex lattice and the singular vortex lattice.
Near the double-core vortex center, there is a sharp low-energy peak, especially along the $a$-axis [Figs.~\ref{LDOS-d}(c) and \ref{LDOS-d}(e)]
and $b$-axis, which is somewhat round [Figs.~\ref{LDOS-d}(d) and \ref{LDOS-d}(f)].
In the double-core case, the OP two minima are situated just outside the center along the $a$-axis [see Fig.~\ref{OP-B}(b)],
giving rise to sharp peaks at a finite energy, as shown in Fig.~\ref{LDOS-d}(e).
On the other hand, the zero-energy peak becomes a bump away from the vortex core for the singular vortex [Figs.~\ref{LDOS-s}(c)-\ref{LDOS-s}(f)].
Note that, in the bulk region away from the vortex core, DOS $\propto|E|^2$ at a low energy, as shown in Figs.~\ref{LDOS-d}(d) and \ref{LDOS-s}(d)
because the gap structure is characterized by two point nodes and two line nodes.
This is also shown in Fig.~\ref{bulkDOS}.

\subsection{NMR spectrum}

\begin{figure}
\begin{center}
\includegraphics[width=8.5cm]{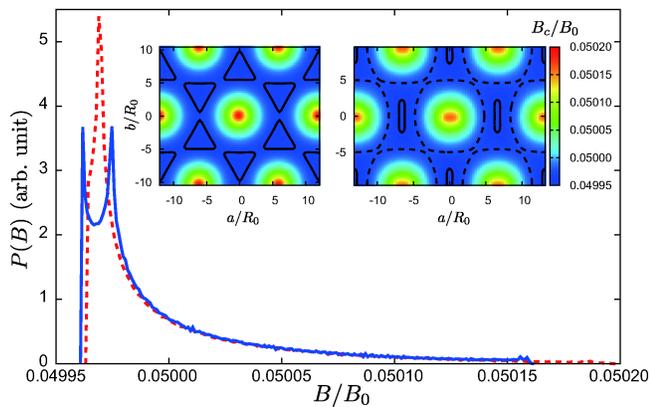}
\end{center}
\caption{(Color online) 
Distribution functions of the internal magnetic field $P(B)$ at $T=0.2T_c$ and $\bar{B}=0.05B_0$
for the singular vortex lattice (dashed line) and double-core vortex lattice (solid line).
The height of $P(B)$ is scaled so that $\int P(B)dB=1$.
Inset: Spatial variations in the internal magnetic field $B_c(\bi{r})$ for the singular vortex lattice (left)
and double-core vortex lattice (right) with the contour lines on the magnetic field situated at the peaks of $P(B)$.
}
\label{NMR-B}
\end{figure}

The double-core vortex lattice is also observed by NMR measurement.
In the NMR experiment, the resonance frequency spectrum of the nuclear spin resonance is determined by the internal magnetic field.
The distribution function is given by
\begin{align}
P(B)=\int\delta[B-B_c(\bi{r})]d\bi{r},
\end{align}
i.e., volume counting for $B$ in a unit cell.
This resonance line shape is called the ``Redfield pattern" of the vortex lattice.
In Fig.~\ref{NMR-B}, we show the distribution functions $P(B)$ for the singular vortex lattice (dashed line) and double-core vortex lattice (solid line).
The distribution function for the singular vortex lattice has a single peak at $B=0.049969B_0$.
This peak comes from the outside of the vortex core, shown by the contour lines in the left inset of Fig.~\ref{NMR-B}.
By contrast, the distribution function for the double-core vortex lattice has a double peak at $B=0.049962B_0$ and $B=0.049975B_0$.
The peaks at the low and high fields come from outside the vortex (solid line) and around the vortex (dashed line), respectively, shown by the contour lines in the right inset of Fig.~\ref{NMR-B}.
The distortion of the double-core vortex lattice gives a clear difference in the NMR spectrum.

\section{C and A Phases}

In the C phase, we take the pairing functions as $\phi=\phi_a=\sqrt{21/8}k_b(5k_c^2-1)$ and $\phi_b=\phi_c=0$,
where the pair potential has one spin component.
Note that the A phase is the same as the C phase except for the exchange between the $a$- and $b$-axes.

\subsection{Morphology of vortex lattice}

In this section, we discuss the deformation of the vortex lattice in the C phase under $H\parallel c$ to determine the effects of the twofold gap function $\bi{d}(\bi{k})\propto\bi{a}k_b(5k_c^2-1)$.
Since the vortex cores are extended along the antinodal $b$-direction,
the height of the triangular lattice is enlarged along the $b$-direction to avoid the overlap of the vortex cores.
This variation of the vortex lattice is also the same for the A phase by rotating it in the $ab$-plane.
To find a stable vortex lattice, we compare the free energies among the triangular lattices with various ratios of the height to the base,
namely, $a_y/a_x$.

\begin{figure}
\begin{center}
\includegraphics[width=7cm]{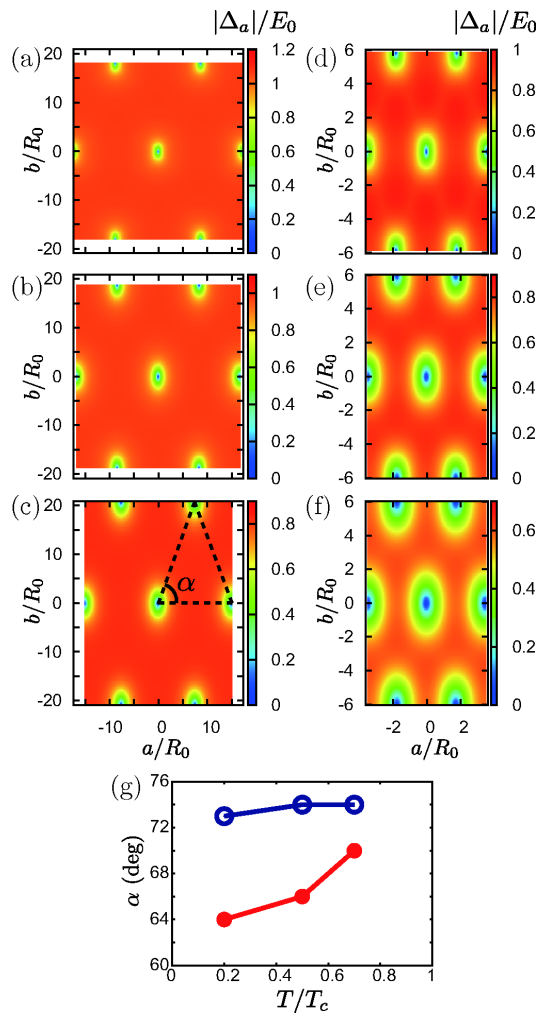}
\end{center}
\caption{(Color online) 
(a)-(f) Spatial variations in the stable pair potential amplitude $|\Delta_a(\bi{r})|$.
(a) $\alpha\approx 64^{\circ }$ at $\bar{B}=0.02B_0$ and $T=0.2T_c$,
(b) $\alpha\approx 66^{\circ }$ at $\bar{B}=0.02B_0$ and $T=0.5T_c$,
(c) $\alpha\approx 70^{\circ }$ at $\bar{B}=0.02B_0$ and $T=0.7T_c$,
(d) $\alpha\approx 73^{\circ }$ at $\bar{B}=0.3B_0$ and $T=0.2T_c$,
(e) $\alpha\approx 74^{\circ }$ at $\bar{B}=0.3B_0$ and $T=0.5T_c$, and
(f) $\alpha\approx 74^{\circ }$ at $\bar{B}=0.3B_0$ and $T=0.7T_c$.
(g) Temperature dependences of the base angle $\alpha$ at $\bar{B}=0.02B_0$ (solid circles)
and $\bar{B}=0.3B_0$ (open circles).
} 
\label{OP-C}
\end{figure}

We show stable vortex lattices at a low magnetic field $\bar{B}=0.02B_0$ in Figs.~\ref{OP-C}(a)-\ref{OP-C}(c)
and at a high magnetic field $\bar{B}=0.3B_0$ in Figs.~\ref{OP-C}(d)-\ref{OP-C}(f).
These figures are also shown at different temperatures, that is, at a low temperature $T=0.2T_c$ in Figs.~\ref{OP-C}(a) and \ref{OP-C}(d),
at an intermediate temperature $T=0.5T_c$ in Figs.~\ref{OP-C}(b) and \ref{OP-C}(e), and
at a high temperature $T=0.7T_c$ in Figs.~\ref{OP-C}(c) and \ref{OP-C}(f).
The triangular lattice is slightly distorted at a low magnetic field and a low temperature, as shown in Fig.~\ref{OP-C}(a).
In this case, the ratio is $a_y/a_x=0.6\sqrt{3}$, that is, 
the base angle of the isosceles triangular lattice is $\alpha\equiv \tan^{-1}(2a_y/a_x)\approx 64^{\circ }$.
As temperature increases, the stable structures of the vortex lattice are $a_y/a_x=0.65\sqrt{3}$, namely, $\alpha\approx 66^{\circ }$ [Fig.~\ref{OP-C}(b)], and $a_y/a_x=0.8\sqrt{3}$, namely, $\alpha\approx 70^{\circ }$ [Fig.~\ref{OP-C}(c)].
The vortex lattice is distorted markedly at a high magnetic field.
At a low temperature, the stable structure is $a_y/a_x=0.95\sqrt{3}$, namely, $\alpha\approx 73^{\circ }$, as shown in Fig.~\ref{OP-C}(d).
Physically, the maximally distorted triangular lattice is $a_y/a_x=\sqrt{17}/2$, namely, $\alpha\approx 76^{\circ }$,
beyond which some of the nearest neighbors are no longer nearest.
Since the distortion of the triangular lattice is near the limit even at a low temperature,
the vortex lattice is hardly distorted by the increase in temperature, as shown in Figs.~\ref{OP-C}(e) and \ref{OP-C}(f),
where $a_y/a_x=\sqrt{3}$, namely, $\alpha\approx 74^{\circ }$.
Thus, the vortex lattice tends to be more distorted to prevent the overlap of the vortex cores at high magnetic fields and at high temperatures
because the ratio of the radius of the vortex core proportional to the coherence length $\xi\propto(T_c-T)^{-1/2}$ 
to the distance between the vortices proportional to $\bar{B}^{-1/2}$ increases.
The deformation of the vortex lattice is summarized in Fig.~\ref{OP-C}(g).

A regular hexagonal vortex lattice was observed in the A phase by the measurement of the SANS~\cite{huxley:2000}.
This result is discussed for the $E_{1g}$ and $E_{2u}$ models theoretically~\cite{champel:2001,agterberg:2002}.
In this experiment, however, since the vortex lattice in the A phase is observed at a low temperature in which the B phase appears with rapid cooling,
the observed vortex lattice may change to the hexagonal singular vortex lattice in the B phase mentioned before.

\subsection{Local density of states}

\begin{figure}
\begin{center}
\includegraphics[width=7cm]{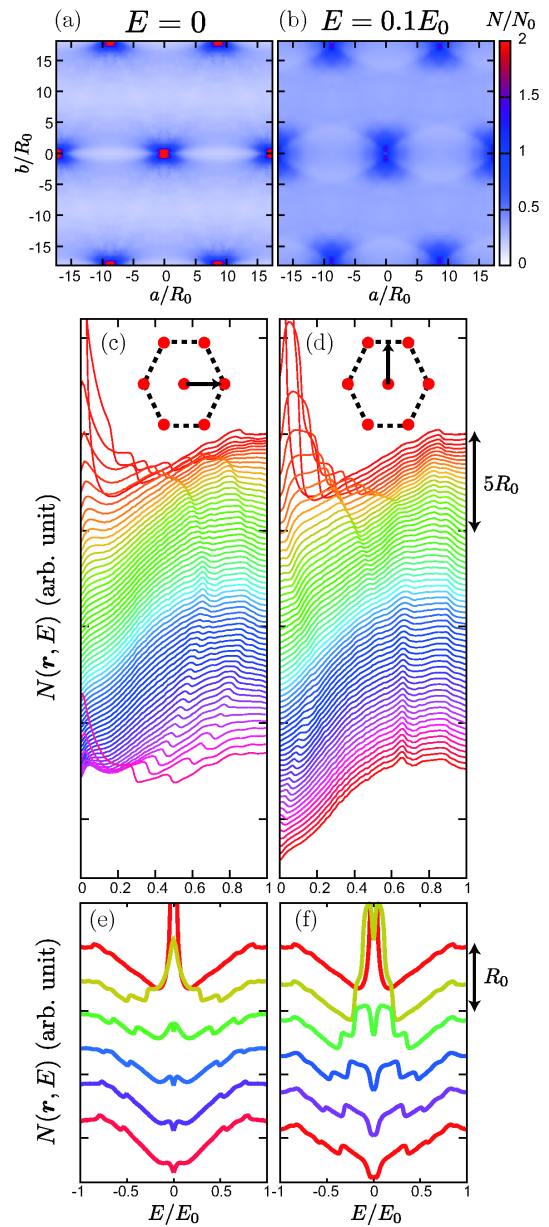}
\end{center}
\caption{(Color online) 
LDOS at $T=0.2T_c$ and $\bar{B}=0.02B_0$ in the C phase.
Spatial variations in the LDOS at $E=0$ (a) and $E=0.1E_0$ (b).
Spectral evolutions of the LDOS from the vortex core (top) along the $a$-axis (c) and $b$-axis (d).
Details of these evolutions near the vortex core are shown in (e) and (f), respectively.
} 
\label{LDOS-C}
\end{figure}

The twofold symmetry of the gap function is revealed, as shown in Fig~\ref{LDOS-C}, where the LDOS is shown.
The zero-energy LDOS [Fig.~\ref{LDOS-C}(a)] is well connected between nearest-neighbor vortices along the $a$-axis, which is the nodal direction.
The LDOS at $E=0.1E_0$ [Fig.~\ref{LDOS-C}(b)] is also well connected between nearest-neighbor vortices;
moreover, it has two peaks within a unit cell aligned along the $b$-axis.
The spectral evolution of the LDOS along the $a$-axis [Figs.~\ref{LDOS-C}(c) and \ref{LDOS-C}(e)] forms a peak structure at a low energy near the vortex core.
At a few $R_0\sim\xi$ away from the vortex core, the spectra still increase from the zero-energy so as to generate a low-energy peak.
On the other hand, it has a rounded bump near the vortex core along the $b$-axis [Figs.~\ref{LDOS-C}(d) and \ref{LDOS-C}(f)].
At a few $R_0$ away from the vortex core, the low energy spectra are almost flat.
Far from the vortex core, their spectra become of the same structure at the center between the vortex cores.
Note that in the bulk region away from the vortex core, the DOS $\propto|E|$ at a low energy, as shown in Fig.~\ref{LDOS-C}(d),
because the gap structure is characterized by one vertical and two horizontal line nodes.
This is also shown in Fig.~\ref{bulkDOS}.

\subsection{Field-angle-resolved zero-energy density of states}

We analyze the field-angle-resolved thermal conductivity experiment~\cite{machida:2012} according to the identified pair function.
It is known that thermal conductivity depends on carrier density and scattering rate, both of which are angle-dependent.
In this experiment, the temperature dependence of thermal conductivity obeys the Wiedemann-Franz law at low temperatures,
implying that QPs play the dominant role in thermal transport.
The most significant effect on the thermal transport in the vortex state comes from the Doppler shift of the QP energy spectrum,
$E(\bi{p})\rightarrow E(\bi{p})-\bi{v}_s\cdot\bi{p}$, 
in the circulating supercurrent flow $\bi{v}_s$.
This effect becomes important at such positions where the gap becomes smaller than the Doppler shift term ($|\Delta|<\bi{v}_s\cdot\bi{p}$).
Thus, we analyze the experimental data~\cite{machida:2012} in terms of field-angle-resolved zero-energy DOS.

Since the magnitude of the Doppler shift strongly depends on the angle between the nodal direction and the magnetic field,
the oscillation of zero-energy DOS occurs.
Consequently, thermal conductivity attains its maximum (minimum) when the magnetic field is directed to the antinodal (nodal) directions~\cite{vekhter:1999,miranovic:2005,sakakibara:2007}.
In this experiment, however, since heat current is injected along the $c$-axis,
thermal conductivity cannot be compared with zero-energy DOS directly.
Then, we compare their differences when field directions are rotated along the vertical line node and antinode in the C phase.

\begin{figure}
\begin{center}
\includegraphics[width=8.5cm]{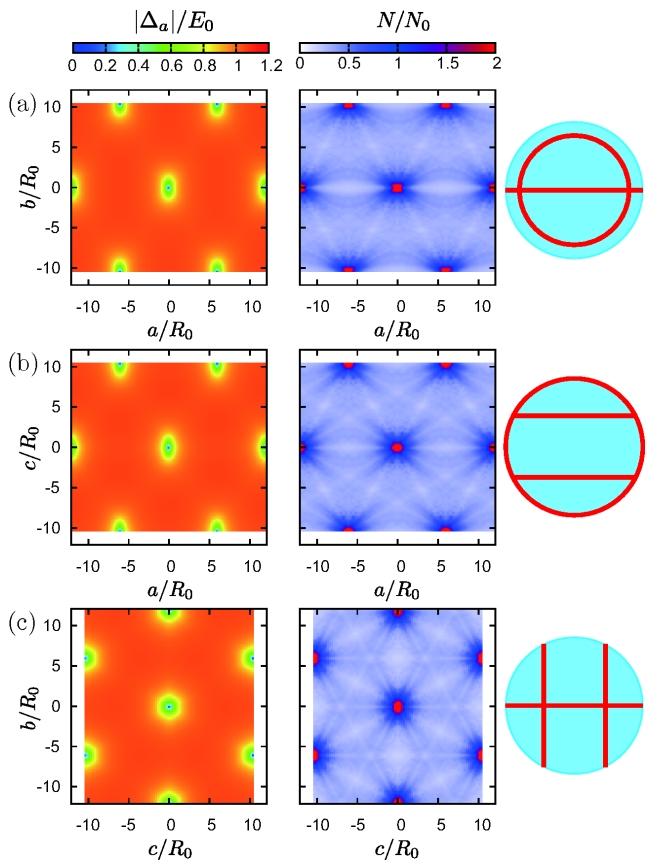}
\end{center}
\caption{(Color online) 
Spatial variations in the pair potential amplitude $|\Delta_a(\bi{r})|$ and
zero-energy LDOS $N(\bi{r},E=0)$ at $T=0.2T_c$ and $\bar{B}=0.05B_0$.
The magnetic fields are directed to the $c$-axis (a), $b$-axis (b), and $a$-axis (c).
Left (middle) panels show the OP amplitude (zero-energy LDOS).
Right panels show a schematic view of line nodes shown in the field direction.
} 
\label{OP-LDOS-Brot}
\end{figure}

In this section, we assume the regular hexagonal vortex lattice $a_y/a_x=\sqrt{3}/2$ or $1/(2\sqrt{3})$.
The stable orientation of the vortex lattice is determined by comparing the free energy calculated using eq.~\eqref{Luttinger-Ward}.
At $T=0.2T_c$ and $\bar{B}=0.05B_0$,
the spatial variations of the pair potential amplitude and the zero-energy LDOS are shown in the left and middle panels of Figs.~\ref{OP-LDOS-Brot}(a)-\ref{OP-LDOS-Brot}(c), respectively.
When the magnetic field is directed to the $c$-axis [Fig.~\ref{OP-LDOS-Brot}(a)] or $b$-axis [Fig.~\ref{OP-LDOS-Brot}(b)],
elliptic vortex cores shrink to the vertical ($H\parallel c$) or tropical ($H\parallel b$) nodal directions. 
Under these magnetic fields, the zero-energy LDOS is mainly connected between nearest-neighbor vortices along the $a$-axis.
On the other hand, under $H\parallel a$ [Fig.~\ref{OP-LDOS-Brot}(c)],
since the vertical node and tropical nodes have similar contributions to the QPs,
vortex cores are hexagonal and the zero-energy LDOS is well connected among all nearest-neighbor vortices,
giving rise to a rather round core profile in Fig.~\ref{OP-LDOS-Brot}(c).

\begin{figure}
\begin{center}
\includegraphics[width=7.5cm]{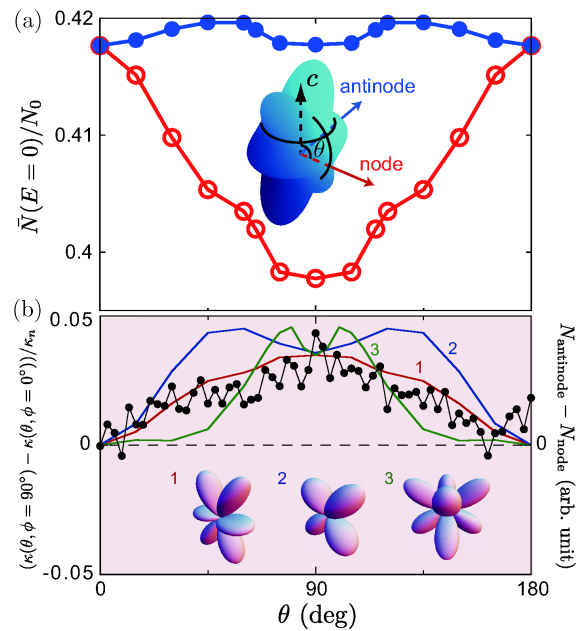}
\end{center}
\caption{(Color online) 
(a) Field-angle-resolved zero-energy DOS at $T=0.2T_c$ and $\bar{B}=0.05B_0$.
The field directions are rotated along the vertical line node (open circles) and antinode (solid circles),
as schematically shown in the inset.
(b) $\theta$ dependence of the thermal conductivity normalized by the normal state value $\kappa_n$ (left axis)
and the DOS differences normalized at $\theta=90^{\circ }$ (right axis)
along the vertical nodal and antinodal scannings for three possible gap functions in the C phase:
1. The present $E_{1u}$ $(k_b(5k_c^2-1))$, 2. $E_{1g}$ $(k_bk_c)$, 3. $E_{2u}$ $(k_ak_bk_c)$.
The gap structures are sketched in the inset.
The experimental data are cited from ref.~\ref{machida2012}.
}
\label{DOS-Brot}
\end{figure}

By taking the spatial average of the zero-energy LDOS's under various field directions, the field-angle-resolved zero-energy DOS is obtained, as shown in Fig.~\ref{DOS-Brot}(a).
When the field direction is rotated along the vertical line node from the $c$-axis (open circles),
the zero-energy DOS is reduced because the number of low-energy excitations from the tropical line nodes decreases.
Within $45^{\circ }<\theta<60^{\circ }$ and $120^{\circ }<\theta<135^{\circ }$, the orientation of the vortex lattice changes,
that is, nearest-neighbor vortices are aligned along the $b$-axis in $60^{\circ }\le\theta\le 120^{\circ }$
and next-nearest-neighbor vortices are aligned along the $b$-axis in $\theta\le 45^{\circ },135^{\circ }\le\theta$.
By contrast, when the magnetic field is rotated along the antinodal direction (solid circles),
the zero-energy DOS is almost constant because the QPs mainly come from the tropical line nodes under $H\parallel c$ and from the vertical line node under $H\parallel b$.
The difference in the zero-energy DOS between the fields along the vertical line node and antinode is maximum at the equator $\theta=90^{\circ }$ because horizontal line nodes are situated in the tropics.
This $\theta$ dependence of the difference in the zero-energy DOS is consistent with the measurement of thermal conductivity~\cite{machida:2012} shown in Fig.~\ref{DOS-Brot}(b).
The earlier spin singlet $d$-wave $E_{1g}$ model ($\phi=\sqrt{15}k_bk_c$) 
and spin triplet $f$-wave $E_{2u}$ model ($\phi=\phi_c=\sqrt{105}k_ak_bk_c,\ \phi_a=\phi_b=0$)~\cite{joynt:2002},
however, have two peaks for the difference in the zero-energy DOS at $\theta\ne 90^{\circ }$ caused by the equatorial line node.

\section{Conclusions and Perspectives}

\begin{table*}[t]
\begin{center}
\caption{Candidate pair functions.}
\begin{tabular}{cccccc}
\hline\hline
irr.~rep. & basis & Knight shift & point + line & twofold in C & gradient coupling \\\hline
$E_{1g}$ & $k_z(k_x,k_y)$ & $\times$ & $\bigcirc$ & $\bigcirc$ & $\times$ \\
$E_{2g}$ & $(k_x^2-k_y^2,k_xk_y)$ & $\times$ & $\times$ & $\times$ & $\triangle$ \\
$E_{1u}^p$ & $\bi{z}(k_x,k_y)$ & $\triangle$ & $\times$ & $\bigcirc$ & $\times$ \\
$E_{2u}$ & $\bi{z}(k_x^2-k_y^2,k_xk_y)k_z$ & $\triangle$ & $\bigcirc$ & $\times$ & $\triangle$ \\
$E_{1u}^f$ & $(\bi{x}k_y,\bi{y}k_x)(5k_z^2-1)$ & $\bigcirc$ & $\bigcirc$ & $\bigcirc$ & $\bigcirc$ \\
\hline\hline 
\end{tabular}
\end{center}
\end{table*}

In this study, we have classified possible pairing functions 
under the given crystalline symmetry $D_{6h}$ for the
heavy-fermion superconductor UPt$_3$, which belong to
the $f$-wave state with a triplet channel.
Then we identified a planar spin triplet state among
them that maximally fits the existing experiments, particularly as follows:

(A) Various bulk thermodynamic measurements that indicate 
the line node(s) and point node(s) in the B phase.

(B) A Knight shift experiment that shows the two field 
directions where the Knight shift decreases below $T_c$,
implying that the $d$-vector contains the $\bi{b}$-component 
and $\bi{c}$-component at lower fields and that
the $d$-vector rotates from $\bi{c}$ to $\bi{a}$
at $H_{\rm rot}$ $(\parallel c)\sim 2$ kG.

(C) An angle-resolved thermal conductivity measurement that
shows a twofold gap structure in the C phase and a rotational symmetry 
in the B phase.

These important experimental results above are all explained by the planar state
with the $f$-wave channel,
namely,  $(\bi{c}k_b+\bi{b}k_a)(5k_c^2-1)$.

In order to check our proposed state, we made several predictions that are
calculated by solving the microscopic Eilenberger equation with our
planar state. The predictions include the following:

(1) The vortex structures in the C and A phases
exhibit a strongly distorted triangular lattice that varies as functions of
$T$ and $H$ when $H\parallel c$. This distortion is caused by the twofold
gap structures in the A and C phases. The vortex morphology should be observed
by SANS experiment.

(2) Although the vortices in the A and C phases are all singular, that is,
the OP vanishes at the core because of the single-component OP,
in the B phase, the vortex is nonsingular, characterized by a double-core structure.
To check this complex vortex structure,
we provide several signatures, such as the magnetic field distribution
probed as the resonance spectrum of  NMR and the LDOS 
around a vortex core probed by STM/STS.

\subsection{Comparison with other proposed states} 

There are several proposals made to identify the pairing symmetries in UPt$_3$.
$E_{1g}$ and $E_{2g}$ are singlet and unable to explain the Knight shift experiment
mentioned above.
$E^p_{1u}$ with the $p$-wave character  is not accepted because there is 
no line node whose existence is firmly established through various thermodynamic 
experiments. $E_{2u}$, which was regarded as the most promising candidate
contradicts the observed twofold gap structure in the C phase.
Table II shows a summary of the present status of various candidates.

\subsection{Remaining issues} 

There remain several issues to be resolved.

(1) $d$-Vector rotation

\noindent 
In order to understand the $d$-vector rotation phenomenon at 
$H_{\rm rot}\sim 2$ kG for $H\parallel c$, we need to take into account the magnetic field energy 
due to the anisotropic susceptibility in the superconducting state.
In the absence of this effect, the induced component $\bi{a}$
near the vortex core center in the double-core phase decreases as the vortex distance decreases.

(2) Origin of symmetry breaking

\noindent
To split $T_c$ into $T_{c1}$ and $T_{c2}$, we need some symmetry breaking field. 
A good candidate is the AF order at $T_N=5$ K observed 
by neutron scattering. This is a high-energy probe for catching
the instantaneous correlation in a snapshot. Other low-energy probes,
such as NMR and $\mu$SR, fail to observe the static AF order.
Since a precise correlation between the AF order and the $T_c$ splitting under
pressure, which simultaneously disappear at $P\sim 3$ kbar, is observed~\cite{trappmann:1991},
this is still puzzling although we previously presented a scenario for this splitting
due to AF fluctuations~\cite{machida:1996}.
An alternative idea is to use the crystal symmetry lowering, which
is also reported before~\cite{midgley:1993,elboussiri:1994,ellman:1995,ellman:cond}.

(3) Pairing mechanism

\noindent
The dipole energy
\begin{align}
H_D\propto\left\langle 3|\bi{k}\cdot\bi{d}(\bi{k})|^2-|\bi{d}(\bi{k})|^2\right\rangle_{\bi{k}}
\end{align}
depends on the combination of the spin and orbital states~\cite{leggett:1975}.
The most favorable combinations by the dipole energy in the $E_{1u}$ state are $\bi{a}k_b(5k_c^2-1)$ and $\bi{b}k_a(5k_c^2-1)$.
In the C phase, the spin state $\bi{b}$ is selected by AF 
ordering and accompanies the orbital state $k_a(5k_c^2-1)$ by a dipole interaction.
In the B phase, the remaining orbital state $k_b(5k_c^2-1)$ has to appear with the spin state $\bi{a}$ to minimize the dipole energy.
However, the pairing state in the B phase without a magnetic field is $(\bi{b}k_a+\bi{c}k_b)(5k_c^2-1)$ actually.
Thus, the combination of the spin and orbital states cannot be interpreted from only the dipole energy.
This special combination between the spin direction and the orbital form hints at the pairing mechanism.
This is one main issue to be resolved in future works since the pairing symmetry is determined in this paper.

(4) Josephson junction experiment

\noindent
By Josephson interferometry experiment, 
the $\pi$ phase shift of the gap function in the B phase for a $90^{\circ }$ rotation about the $c$-axis was proposed~\cite{strand:2009}.
Also, by measuring the critical current on Josephson tunnel junctions between the $a$- and $b$-axes, 
the nodal direction of the gap function in the C phase was proposed at $45^{\circ }$ with respect to the $a$-axis~\cite{strand:2010}.
Their conclusions are consistent not with our $E_{1u}$ model but with the $E_{2u}$ model.

(5) Topological aspect

\noindent
The identified pairing state $(\bi{c}k_b+\bi{b}k_a)(5k_c^2-1)$ is analogous to
the superfluid $^3$He B-phase whose form is described by $\bi{x}k_x+\bi{y}k_y+\bi{z}k_z$ 
realized in the bulk or to the planar state $\bi{x}k_x+\bi{y}k_y$ realized in thin films.
Our state has Majorana particles at the boundary as an Andreev bound state,
albeit line and point nodes exist in the bulk.
This is interesting because recently Sato has argued the possibility of topological
protection under a nodal gap~\cite{sato:private}.
The topological nature is discussed in a similar situation 
in connection with the superfluid $^3$He A-phase where
Majorana particles exist in a point node gap\cite{tsutsumi:2010b,tsutsumi:2011b}.
This topological aspect certainly deserves further investigation.
Note that the double-core vortex does not contain the Majorana zero mode.

\section*{Acknowledgments}

We thank K.~Izawa, Y.~Machida, T.~Sakakibara, and S.~Kittaka for informative discussions on their experiments
and M.~Ichioka for formulations and computational coding.
This work was supported by JSPS, KAKENHI (No.~21340103).
Y.T. acknowledges the financial support from the JSPS Research Fellowships for Young Scientists.


\end{document}